\documentclass[pre,twocolumn,superscriptaddress,preprintnumbers,amsmath,amssymb,showpacs,showkeys]{revtex4}
\usepackage{graphicx}
\usepackage{dcolumn}
\usepackage{bm}
\usepackage{epsfig}
\usepackage[dvipsnames]{color}

\newcommand{\beq}{\begin{eqnarray}}
\newcommand{\eeq}{\end{eqnarray}}
\newcommand{\Tr}{{\text Tr}}
\newcommand{\e}{{\text e}}
\newcommand{\rmd}{{\text d}}

\def\bP{ \boldmath P}
\def\ba{{\mbox{\boldmath $\alpha$}}}
\def\bb{{\mbox{\boldmath $\beta$}}}

\def\bchi{{\mbox{\boldmath $\chi$}}}
\def\bEta{{\mbox{\boldmath $\eta$}}}
\newcommand{\figurewidth}{\columnwidth}
\input xy 
\xyoption{all}
\definecolor{myback}{rgb}{1,.964,.8}

\begin{document}

\title{Excitation Gap from Optimized Correlation Functions \\ in Quantum Monte Carlo Simulations}
\author{Itay ~Hen}
\email{itayhe@physics.ucsc.edu}
\affiliation{Department of Physics, University of California, Santa Cruz, California 95064, USA}

\date{\today}

\begin{abstract}
We give a prescription for finding optimized correlation functions for the extraction of the gap to the first excited state within quantum Monte Carlo simulations. 
We demonstrate that optimized correlation functions 
provide a more accurate reading of the gap when compared 
to other `non-optimized' correlation functions and are generally characterized by
considerably larger signal-to-noise ratios. 
We also analyze the cost of the procedure and show 
that it is not computationally demanding. 
We illustrate the effectiveness of the proposed procedure by analyzing several
exemplary many-body systems of interacting spin-$1/2$ particles.   
\end{abstract}

\pacs{02.70.Ss,03.67.Lx,05.10.Ln} 

\keywords{Excitation gap, Quantum Monte Carlo simulations, Imaginary time correlation functions} 
\maketitle

\section{Introduction}
Quantum Monte Carlo (QMC) simulations have become
the method of choice for studying large equilibrium quantum
many-body systems without approximations in more than one dimension  \cite{qmcRef1,qmcRef2} (in one dimension, the Density Matrix Renormalization Group has proven to be an extremely powerful tool \cite{DMRG1,DMRG2}).
While for small system sizes 
one may employ exact-diagonalization techniques, for larger ones, QMC methods provide in most cases the only numerical method available for exact numerical 
investigation.
However, employing Monte Carlo techniques comes at a cost. As the name itself might indicate, QMC methods are stochastic in nature
as they are based on sampling the exponential number of states of the Hilbert space of the system, and there are therefore statistical errors associated with every measured quantity.

QMC methods are usually considered ideal for measurements of ground-state properties or for the determination 
of thermodynamic properties of physical systems, as these can usually be measured to a high degree of accuracy, i.e., with very small statistical errors. 
While this is true, QMC methods are also
known to be less suited for extracting information about excited states, which tend to be rather cumbersome to obtain and are typically measured with much less accuracy.

Excited states play a central role in many areas of physics and chemistry of many-body systems.
Among these are critical phenomena and phase transitions in Condensed Matter Physics \cite{excitedRefPhys}, mass gap calculations
in High Energy Physics \cite{massgap1,massgap2,massgap3},
various calculations pertaining to the properties of nuclei in Nuclear Physics \cite{excitedNuc,excitedNuc2} and the vibrational modes of large molecules in Chemistry \cite{excitedRefChem,excitedRefChem2}, to mention
some diverse examples. 
Naturally, numerous attempts have been made to utilize quantum Monte Carlo methods to compute excited-state energies as well.
However, this has turned out to be a difficult task because obtaining information about excited states involves `isolating' specific regions within the spectrum of the Hamiltonian --
something which can not be done by simple measurements of thermodynamic properties (except maybe the ground state at ultra-low temperatures).

As quantum Monte Carlo methods have evolved over the years, techniques to extract information about excited states have been continuously developed.
Most of these were based on some form of analysis performed on measurements of imaginary-time correlation functions as these provide indirect access 
to the spectrum of the system Hamiltonian \cite{maxEnt,maxEnt2,maxEnt3} (this will be explained in more detail in the next section). 
Despite these elaborate manipulations on the measured data, an accurate prediction of even the lowest excitation energies still remains a challenge -- and under some circumstances an impossibility -- due to the large statistical errors associated with correlations with large imaginary-time differences,
although methods to reduce these errors in certain cases through the use of improved estimators in cluster-based QMC methods
\cite{improved} or the use of 	`smeared' operators in lattice gauge theories \cite{smeared1,smeared2} have been devised.

In what follows, we propose a remedy to these difficulties
by suggesting a way to partially optimize the manner in which excited-state energies, specifically the gap to the first excited state, are calculated from imaginary-time correlation functions. We do
this by providing a prescription to find and then measure the most suitable correlation function available for this purpose (within some stated limitations).
The method we suggest here is based on finding the operator whose imaginary-time correlation function is optimal for the extraction of excited-state energies, 
where the optimization is based on the maximization of the integrated susceptibility within a space of `basic operators' and under appropriate constraints. 
As we shall demonstrate, this type of optimization removes, or at least substantially reduces, 
some of the difficulties associated with dealing with the large statistical errors that characterize correlations with large imaginary-time separations.

The paper is organized as follows. In Sec.~\ref{sec:ac} we discuss in some detail
the basics of extracting excited-state energies from imaginary-time correlation functions, focusing on the extraction of the gap to the first excited state. We also list some of the difficulties involved in doing so. 
In Sec.~\ref{sec:ocf} we shall present a method to find a measurable operator whose imaginary-time correlation function is
optimal for the extraction of the gap. We provide several illustrative examples of the method in Sec.~\ref{sec:example} and 
summarize the results in Sec.~\ref{sec:sum} along with some conclusions. 

\section{\label{sec:ac}Accessing the excitation gap -- correlation functions in imaginary time}
Let us consider a many-body system described by the Hamiltonian $\hat{H}$
(imagine, say, an $N$-body system of interacting spin-$1/2$ particles) 
at inverse-temperature $\beta=1/T$ (in our units, $k_B=1$). 
The thermal averages of physical observables are given by
\footnote{For the discussion here we shall consider the canonical ensemble scheme, although the following may just as well be applied to the grand-canonical ensemble.}:
\beq
\langle\hat{O} \rangle=\frac1{Z} \times \Tr[\hat{O} \e^{-\beta \hat{H}}] \,,
\eeq
where $\hat{O}$ is the operator associated with the physical observable in question. 
Here, $Z$ is the partition function $Z=\Tr[ \e^{-\beta \hat{H}}]$ and $\Tr$ is the trace operation.

If the system is small enough, these thermal averages may be computed rather easily
by exact-diagonalization techniques, with which the full matrix of the Hamiltonian is spectrally decomposed. 
In this case, excited-state energies would simply be 
obtained by subtracting the lowest eigenvalue of the Hamiltonian matrix from other eigenvalues. 

For bigger systems however, where exact-diagonalization methods are unfeasible, one must almost always resort to QMC techniques to obtain accurate results. 
While the thermal averages of most operators of potential interest are usually very easy to obtain within QMC simulations (as discussed in the Introduction),
in order to evaluate excited-state energies of the system, only somewhat-indirect methods are 
available.

The first step toward finding excited-state energies is the calculation of the thermal averages of different-time correlations 
of measurable operators that do not commute with the Hamiltonian (and hence are not conserved in time).
Consider the imaginary-time `two-point' correlation of the measurable operator $\hat{O}$, namely $\langle \hat{O}(\tau) \hat{O}(0) \rangle$,
where $\tau$ is the imaginary-time coordinate. This expression may be expanded in the eigen-energy basis to give:
\beq \label{eq:oto}
\langle \hat{O}(\tau) \hat{O}(0) \rangle &=& 
\langle \e^{\hat{H} \tau} \hat{O}(0) \e^{-\hat{H} \tau} \hat{O}(0) \rangle = \\\nonumber 
\left( \sum_{k=0} \e^{-\beta E_k}  \right)^{-1} &\times&  \sum_{n,m=0} |\langle n | \hat{O} | m \rangle |^2 \e^{-(E_m-E_n) \tau} \e^{-\beta E_n} \,,
\eeq
where $\{ E_n \}$ and $\{ | n\rangle \}$ are the eigenvalues and matching eigenstates of the Hamiltonian $\hat{H}$. 
Now, if $\beta$ is chosen such that $\beta \Delta E_1 \gg 1$ where $\Delta E_n=E_n-E_0$ is the gap to the $n$-th excited state (and in particular the excitation gap is $\Delta E_1$), it is expected that the system will eventually relax to its ground state $|0\rangle$ \footnote{Of course, in some cases where critical slowing down is unavoidable, especially in the vicinity of first order phase transitions, relaxation
to the ground state may take an exponentially-long amount of time.}.
Under this condition, which we shall assume to hold henceforth, one could define the following correlation function with the associated series expansion:
\beq
\label{eq:corrFunc}
C_{\hat{O}}(\tau) &\equiv& \langle \hat{O}(\tau) \hat{O}(0) \rangle- \langle \hat{O} \rangle ^2  \\\nonumber
&\approx&
\sum_{n=1} |\langle 0 | \hat{O} | n \rangle |^2 \left( \e^{-\Delta E_n \tau} 
+\e^{-\Delta E_n (\beta- \tau)}\right) \,.
\eeq

Information about excited-state energies (or equivalently the gaps $\Delta E_n$)
is usually extracted by fitting measurement data of the correlation function or some transformation thereof, to an expression similar to the sum in the above equation,
 where usually the free parameters of such fits correspond to the energy gaps $\Delta E_n$ and the matrix elements $|\langle 0 | \hat{O} | n \rangle |$. 
For obvious reasons, finding more than a few excited-state gaps 
is unfeasible because of the exponential number of free parameters involved in the fit
and the finite number of available uncorrelated measurement data points
(although attempts to model the some of the spectrum by a continuum of states has also been suggested \cite{cont}). 

Here, we shall focus the discussion on a rather basic type of analysis of the imaginary-time correlation function data with which the gap to the first excited state is extracted,
although it should be noted that
more sophisticated methods of analysis exist and may be employed just as easily. 
In fact, these methods are expected in most cases to
perform better than the simple analysis and provide better estimates of the excitation gap and possibly also limited information on higher energy levels \cite{maxEnt,maxEnt2,maxEnt3}. However, application of these methods of analysis and comparison between them is complementary to the discussion here 
and will therefore remain outside the scope of this paper. 

Examination of the form of the correlation function given in Eq.~(\ref{eq:corrFunc})
suggests that it might be possible to extract the excitation gap by analyzing the behavior of the correlation function
at long imaginary times
where the slowest-decaying exponent dominates the series and as a result the correlation function may be approximated 
in that region by:
\beq
C_{\hat{O}}(\tau) \approx |\langle 0 | \hat{O} | 1 \rangle |^2 \e^{-\Delta E_1 \tau}\,.
\eeq
In this case, the simplest and most straightforward method of analysis for extracting
the gap $\Delta E_1$ would simply be fitting the logarithm of the obtained measurement data of $C_{\hat{O}}(\tau)$ acquired in the simulation with 
a straight line in the said region.
For the gap to be obtained accurately, however,
in addition to being dominated by the slowest-decaying term, the correlation
function must also have small relative statistical errors, i.e., 
 a large signal-to-noise ratio  -- which 
is usually a feature of short imaginary-time correlations. 
Normally then, one looks for an intermediate region between $\tau=0$ and $\tau=\beta/2$ (the correlation function is symmetric about $\beta/2$)
that satisfies both of the above demands. It should also be noted that the choice of $\beta$ also plays a role in finding an appropriate region: If $\beta$ is chosen to be too small,
there will be no region where only the slowest-decaying exponent survives. On the other hand, if $\beta$ is chosen to be too big, the system will take longer 
to equilibrate.

\begin{figure}
\begin{center}
\includegraphics[width=\figurewidth]{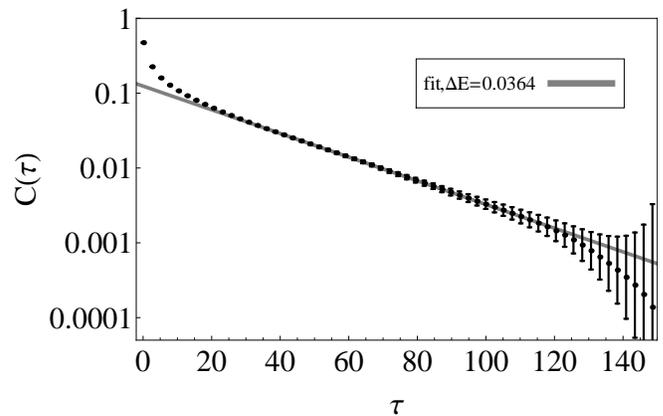}
\caption{A log-linear plot of a time dependent correlation function for an $N = 64$ spins system of one instance of the 1-in-3SAT problem with $\beta=1024$
(further details of the problem can be found in Sec.~\ref{sec:example}). 
While at early times there are contributions from several energy levels, there is a region where the correlation function 
may be fit with a simple straight line (on a log-linear scale) from which the energy gap which is the negative of the slope could be obtained (giving here $\Delta E_1=0.0364$).
At still later times where the correlation function is significantly attenuated, the statistical errors become huge. }
\label{fig:ctau}
\vspace{-0.7cm}
\end{center}
\end{figure}

An illustrative example of the method is given in Fig.~\ref{fig:ctau}. In the figure, results of the above analysis applied to a system of $64$ interacting spin-$1/2$ particles encoding a 64-bit 1-in-3SAT problem augmented by transverse fields 
is presented. (The specific structure of the Hamiltonian of the system is discussed in Sec.~\ref{sec:example}.)
The figure shows the correlation function of the diagonal part of the Hamiltonian 
of the system accompanied by a linear fit in an intermediate region
aimed at evaluating the gap of the system. 

As the figure indicates, choosing the appropriate region in imaginary time is not always a simple task: At small $\tau$, all the exponents in the sum given in Eq.~(\ref{eq:corrFunc}) 
are expected to contribute significantly to $C_{\hat{O}}(\tau)$
in proportion to the square of their respective matrix elements $|\langle 0 | \hat{O} | n \rangle | ^2$. 
At the other end of the range, at times closer to $\beta/2$, it is very likely that 
the slowest-decaying exponent will be the only surviving one (provided that $\beta$ is large enough), 
however in practice, it may also be very difficult to obtain an accurate estimate of the correlation function 
there, since the signal-to-noise ratio of correlations with significant time differences is typically very small due to the exponential decay of the function
 (this is also evident in Fig.~\ref{fig:ctau}). 

Possible improvements over the above method of analysis would involve
for example, fitting the correlation function with hyperbolic cosines which would account for the signal coming from $\beta/2< \tau < \beta$,
or adjusting the fit to partially account for contributions of higher-energy excitations. These methods will not be discussed further here. 

\section{\label{sec:ocf}Optimized correlation functions}
In the previous section, we saw that our capability of accurately extracting the gap depends on
the existence of a region where the correlation function can be nicely fit with a straight line (on a log-linear scale).
The expression for the correlation function given in Eq.~(\ref{eq:corrFunc}) 
shows that for a given value of $\beta$, the exact shape of the correlation function is dependent upon two sets of values. The first is the spectrum of $\hat{H}$ which is a given property 
of the system. The second set of values is the matrix elements $| \langle 0 | \hat{O} | n \rangle|^2$ which on
the other hand can be partially manipulated by different choices of the operator $\hat{O}$ whose different-time correlations are being measured. 
The general guideline for choosing the most suitable operator to measure 
the correlation function of, for the extraction of the gap would naturally be based on bringing
the matrix element $| \langle 0 | \hat{O} | 1 \rangle|^2$ to a maximum while keeping the other matrix elements $| \langle 0 | \hat{O} | n \rangle|^2$ for $n>1$ constrained.
Doing so would have a two-pronged effect on the correlation function: 
it would yield a more dominant slowest-decaying exponent at short times and will also make the correlations stronger throughout. This would in turn substantially reduce the statistical errors at intermediate and long times.

\subsection{\label{sec:maxSus}Maximizing the integrated susceptibility}

The matrix elements $| \langle 0 | \hat{O} | n \rangle|$ discussed above can not however be accessed or manipulated directly within
quantum Monte Carlo simulations as this requires knowledge of the excited states of the system -- knowledge that one does not have
when the system is in its ground state.
 
A question then arises as to how one could determine which measurable operator 
$\hat{O}$ has the optimal imaginary-time correlation function
for the extraction of the excitation gap?
As it turns out, there is a measurable thermodynamic physical quantity associated 
with every operator that may provide indirect access to these matrix elements
and as we shall see will prove to be key in finding a well-suited operator for the extraction of the gap \cite{sandvik}. This quantity is the `integrated susceptibility' of the operator which is defined by:
\beq
\chi_{\hat{O}} &\equiv& \int_{0}^{\beta} C_{\hat{O}}(\tau) \rmd \tau \\\nonumber
&\approx&
\sum_{n=1} \frac{|\langle 0 | \hat{O} | n \rangle |^2}{\Delta E_n} \times 2(1-\e^{-\Delta E_n \beta}) \,.
\eeq
The above quantity has two very favorable properties. Firstly, it is a zero-frequency quantity,
and while it involves integration over the entire range of imaginary time, it can still be very easily and efficiently measured in the course of a simulation within a variety of QMC algorithms (see Appendix~\ref{app:sseImp} for a description of its measurement within the Stochastic Series Expansion algorithm). Secondly, as the sum in the above equation indicates, 
the integrated susceptibility may be viewed as an estimator
or measure of the (squared) matrix element to the first excited state $|\langle 0 | \hat{O} | 1\rangle|^2$:
In particular, in cases where the second and higher excited states have considerably higher energies than that of the first excited state, $\chi_{\hat{O}}$ could be approximated by
\beq
\chi_{\hat{O}} \approx 2 \frac{|\langle 0 | \hat{O} | 1 \rangle |^2}{\Delta E_1} \,,
\eeq
and one could therefore use the integrated susceptibility as an indication for the magnitude of $\langle 0 | \hat{O} | 1 \rangle |^2$.
The integrated susceptibility may thus be used as a `figure of merit'
for the effectiveness of any candidate $C_{\hat{O}}(\tau)$:
 the larger $\chi_{\hat{O}}$ is, the
better $C_{\hat{O}}(\tau)$ would be for extracting the gap. 
Graphically, maximizing $\chi_{\hat{O}}$ corresponds to `lifting' 
the correlation function curve as much as possible above the horizontal axis
thereby maximizing the area underneath in the region $[0,\beta]$.

Given the above discussion, we are now at a point where we can reformulate in a more concrete manner the question of finding the optimal 
correlation function
$C_{\hat{O}}(\tau) = \langle \hat{O}(\tau) \hat{O}(0) \rangle- \langle \hat{O} \rangle^2$
for the extraction of the excitation gap: 
Suppose that 
within a QMC simulation, there exists a set of $M$ basic observables 
$\{ \hat{A}_1,\ldots,\hat{A}_i,\ldots,\hat{A}_M \}$ which can be easily measured
in the course of the simulation \footnote{Obviously, one could choose to work with only a subset of the available operators or rather with
combinations of them, thereby reducing the total number of operators $M$.}. 
What would be the operator $\hat{O} = \sum_{i=1}^M \alpha_i  \hat{A}_i$, 
where $\alpha_i $ are real-valued coefficients, such that $\chi_{\hat{O}}$ is maximal?

Expressing the integrated susceptibility of the operator $\hat{O}$ in terms of the coefficients $\alpha_i$, we have:
\beq
\chi_{\hat{O}} = \sum_{ij} \alpha_i \alpha_j \chi_{ij} \,,
\eeq
where
\beq
\label{eq:chiij}
\chi_{ij} = \int_0^{\beta} C_{ij}(\tau) \rmd \tau\,.
\eeq
Here, the `basic' correlation functions $C_{ij}(\tau)$ are defined as:
\beq
&&C_{ij}(\tau)=\\\nonumber
&&\frac1{2} \left( \langle \hat{A}_i(\tau) \hat{A}_j(0)\rangle + \langle \hat{A}_j(\tau) \hat{A}_i(0)\rangle\right) - \langle \hat{A}_i(0)\rangle \langle \hat{A}_j(0)\rangle \,.
\eeq

Next, we note that multiplication of the correlation function $C_{\hat{O}}(\tau)$ by an arbitrary constant factor simply corresponds to multiplying 
the operator $\hat{O}$ by the square of that constant, and we should therefore restrict the discussion to normalized correlation functions. 
This is done by the natural requirement that the value of the correlation function at $\tau=0$, namely $C_{\hat{O}}(0)$, be one.
In terms of the coefficients $\alpha_i$, this translates to the condition:
\beq
\sum_{i j} \alpha_i \alpha_j \eta_{ij}= 1 \,,
\eeq
where 
\beq
\label{eq:etaij}
\eta_{ij}=C_{ij}(0)=\frac1{2} \left( \langle \hat{A}_i \hat{A}_j \rangle + \langle \hat{A}_j \hat{A}_i \rangle \right) -\langle \hat{A}_i \rangle \langle \hat{A}_j \rangle \,.
\eeq
Note that unlike the matrix elements $\chi_{ij}$ which depend on the long-time behavior of the system, i.e., on the entire spectrum of the Hamiltonian,
the matrix elements $\eta_{ij}$ are ground-state properties.

Interestingly, the above normalization condition can be expressed as an inner product 
in this `space of basic operators' $\{ \hat{A}_i \}$. 
For any two arbitrary measurable operators $\hat{A}$ and $\hat{B}$, this inner product is simply the equal-time covariance:
\beq
\hat{A}*\hat{B} \equiv \frac1{2} \left( \langle  \hat{A} \hat{B} \rangle +\langle \hat{B} \hat{A} \rangle \right) 
-\langle \hat{A}\rangle \langle \hat{B} \rangle  \,,
\eeq
from which the norm 
\beq
|| \hat{A} ||  \equiv \left(\hat{A}*\hat{A}\right)^{1/2} = \left(\langle \hat{A}^2 \rangle -\langle \hat{A}\rangle^2 \right)^{1/2} 
\eeq
is immediately derived.

Denoting the vector of the coefficients of $\hat{O}$ by \hbox{$\ba=(\alpha_1,\ldots,\alpha_i,\ldots,\alpha_M)$},
our problem translates to maximizing the quantity
$\langle \ba | \bchi | \ba \rangle$, where $ \bchi $ is the positive definite matrix whose $ij$-th entries are $\chi_{ij}$,
supplemented by the normalization condition $||\hat{O}||=1$ which translates to
\beq
\langle \ba | \bEta | \ba \rangle =1 \,,
\eeq
where $\bEta$ is the equal-time covariance matrix whose entries are, analogously, $\eta_{ij}$. It too is a positive definite matrix.
Both matrices $\bEta$ and $\bchi$ fall under the category of Gramanian matrices for which each entry
can be viewed as an inner product of two elements chosen from a set of $M$
elements. As we shall see later, the two sets of entries, $\chi_{ij}$ and $\eta_{ij}$ have something else is common: 
both can be very easily measured within QMC simulations.

\subsection{Projecting out conserved quantities}
Before moving on to the maximization procedure however, there is another delicate issue that needs to be addressed:
It may very well be the case that the Hamiltonian governing the physical system in question has a set of conserved quantities associated with it
(one of whom would usually be the Hamiltonian itself). Let us denote the operators corresponding to these quantities by the
set $\{ \hat{B}_k \}$ with $k=1..N_c$ 
where  $N_c$ is the `number of linearly independent constraints' or equivalently the `number of conserved quantities', 
and rewrite each of those if possible 
as a linear combination of the set of the basic operators $\{ \hat{A}_i \}$ that comprise $\hat{O}$ 
\footnote{It should be noted that in principle there could be a situation where a conserved quantity could not be constructed 
from the set of basic operators.}:
\beq
\hat{B}_k=\sum \beta^{(k)}_i \hat{A}_i \,,
\eeq
where $\beta^{(k)}_i$ are real-valued coefficients. 
The normalized correlation functions corresponding to these conserved quantities are all simply $C_{\hat{B}_k}(\tau)=1$, i.e., they 
are constant in imaginary time.

It is important to note then that maximizing $\chi_{\hat{O}}$ without taking these operators into account is guaranteed to produce 
the `optimized' correlation function $C_{\hat{O}}(\tau)=1$,
with the maximal value of $\chi_{\hat{O}}=\beta$
corresponding to an operator which is an arbitrary linear combination of these conserved quantities. 
This of course is a situation that one would wish to avoid, as the gap could not be extracted
from a constant correlation function.
It is therefore necessary to `remove' the above conserved quantities from the optimized operator $\hat{O}$. 
Interestingly, in terms of the newly defined inner product, this condition may be formulated very naturally by requiring that $\hat{O}$ be orthogonal to, or in other words uncorrelated with, each of the operators corresponding to the various conserved quantities,
namely by requiring that $\hat{O}*\hat{B}_k= 0$ for each $k$.
In vector notation, these requirements translate to:
\beq
\label{eq:ortCons}
\langle \ba | \bEta | \bb^{(k)} \rangle = 0 \,.
\eeq

The careful reader will notice that there is a certain subtlety associated with the condition Eq.~(\ref{eq:ortCons}), which is important to address. 
As discussed earlier, it is crucial for the extraction of the gap that our system be strictly in its ground state,
as we shall assume it is for all practical purposes. 
In the ground state, the conserved quantities $\{ \hat{B}_k \}$ do not fluctuate. They obey
\beq
\langle \hat{B}_k^2\rangle -\langle \hat{B}_k\rangle^2=0 \,.
\eeq
In terms of the equal-time covariance matrix $\bEta$ (and also in terms of the matrix $\bchi$) this translates to $\langle \bb^{(k)} | \bEta | \bb^{(k)} \rangle =0$. 
The above equation implies that $\bEta$ is in practice no longer strictly positive definite but only positive {\it semidefinite} and the set of
vectors $ \{ \bb^{(k)}\}$ spans its kernel (and also that of $\bchi$). 
In this case, it would be impossible to require that the condition Eq.~(\ref{eq:ortCons}) be satisfied. 
For the optimized operator $\hat{O}$ to be orthogonal to those, we must restrict $\hat{O}$ to the subspace orthogonal to the kernel of $\bEta$,
that is, to require that the vector $\ba$ satisfies the amended conditions
\beq
\langle \ba | \bb^{(k)} \rangle =0 \,,
\eeq
for all $k=1..N_c$. 

In passing, we note the following: It may very well be the case that one would not be aware of all the conserved quantities 
associated with a given Hamiltonian and that can be given as linear combinations of the basic operators.
Therefore, in the course of optimizing the correlation function for the extraction of the gap,
the to-be-optimized operator $\hat{O}$ will not be orthogonal to all conserved quantities. In this case, the maximization of $\chi_{\hat{O}}$
will yield some linear combination of operators from the set $\{ \hat{A}_i\}$ which would again give $C_{\hat{O}}=1$. 
The resulting operator will produce a correlation function with which one could not extract the gap, as the latter would just be a constant.
The procedure would however yield the unknown conserved quantity.
Put differently, the maximization process described above may be used to { \it detect } unknown conserved quantities of the system 
(although this would probably require very accurate measurements and may therefore turn out to be a numerically very demanding task). 
Once all the conserved quantities are revealed, they may then be removed from $\hat{O}$ by suitable orthogonality conditions.   

\subsection{\label{sec:optProc}The optimization process}
At this stage, we can reformulate the problem at hand in a purely mathematical form:
Given two symmetric $M \times M$ square matrices $\bchi$ and $\bEta$ and a set of vectors $(\bb^{(1)}, \ldots, \bb^{(k)}, \ldots, \bb^{(N_c)})$
with $N_c<M$, find the vector $\ba $ that maximizes
$\langle \ba | \bchi | \ba \rangle$ given the constraints $\langle \ba | \bEta | \ba \rangle=1$
and  $\langle \ba | \bb^{(k)} \rangle=0$ for all $k=1..N_c$. 

The solution to the above problem is easily obtainable and is given by the following prescription:
\begin{itemize}
\item
Using a Gram-Schmidt orthonormalization process, find a set of $(M-N_c)$ orthonormal vectors $\{ \bb_{\perp}^{(k)}\}$ with $k=1..M-N_c$, that span the subspace orthogonal to that
spanned by the set $\{\bb^{(k)}\} $. Construct then the $(M-N_c) \times M$ matrix ${\bf \bP_{\perp}}$ whose rows are the vectors $\{ \bb_{\perp}^{(k)}\}$. 
\item
Define the `reduced' matrices $\bar{\bEta}= \bP_{\perp} \bEta \bP^{\dagger}_{\perp}$ and $\bar{\bchi}=\bP_{\perp} \bchi \bP^{\dagger}_{\perp}$.
These are just $\bEta$ and $\bchi$ with the subspace spanned by the set $\{\bb^{(k)}\}$ removed. 
In our case, this
ensures that  $\bar{\bEta}$ is a positive definite matrix. The dimensionality of each of the new matrices is $(M-N_c) \times (M-N_c)$.
\item
Now the problem reduces
to finding a vector $ \bar{\ba} $ that maximizes:
\beq
\label{eq:maxExpression}
\frac{ \langle  \bar{\ba} | \bar{\bchi} |  \bar{\ba} \rangle} {\langle  \bar{\ba} | \bar{\bEta} |  \bar{\ba} \rangle} \,.
\eeq
The above expression would be maximal if we set $| \bar{\ba} \rangle$ to be the eigenvector of $\bar{\bEta}^{-1/2} \bar{\bchi} \bar{\bEta}^{-1/2} $ belonging to the largest eigenvalue.
\item 
Switching back to the full Hilbert space, the vector
\beq
\label{eq:sol}
| \ba \rangle =\bP^{\dagger}_{\perp}  \bar{\bEta}^{-1/2} | \bar{\ba} \rangle
\eeq
is then the solution to our problem.
\end{itemize}
 \subsection{Practical guidelines and cost of the procedure}
Taking time-correlation measurements of the optimized composite operator $\hat{O} = \sum_{i=1}^M \alpha_i \hat{A}_i$
using the optimal set $\{ \alpha_i \}$, requires first knowing the values of these coefficients up to an acceptable statistical error. 
In the previous section, we saw that the values of these coefficients are given as functions of the matrix elements $\chi_{ij}$ and $\eta_{ij}$
and these need to be determined beforehand. Thus, measuring the imaginary-time correlations of $\hat{O}$ requires 
that the QMC simulation has two `phases' as far the gap calculations are concerned: 
In the first phase, the optimal set $\{ \alpha_i \}$ needs to be determined by performing measurements of $\chi_{ij}$ and $\eta_{ij}$ until the desired accuracy is reached. 
At the end of this phase, the
set of parameters $\{ \alpha_i \}$ is calculated according to the prescription given in Sec.~\ref{sec:optProc}.
Thus, during the first phase of the simulation no imaginary-time correlations are measured; the actual measurements of the correlation function corresponding to the optimal operator are performed in a second phase of the simulation.

Measurements of $M^2$ physical quantities during the first phase of the simulation may seem a bit costly at first
due to the fact that the number of independent `basic' operators in a given problem usually scales like the number of particles in the system $N$, that is, $M^2 \sim N^2$. 
However, we note that this seemingly high cost is compensated by the fact that the 
number of operations needed for this measurement process 
does not scale with, and in fact is independent of, the inverse temperature $\beta$.
For ground-state measurements, the appropriate inverse-temperature $\beta$ 
normally grows polynomially or even exponentially with $N$.
This is because of the condition of $\beta \Delta E_1 \gg 1$ which needs 
to be maintained while the gap to the first excited state $\Delta E_1$ usually decreases at least 
polynomially fast in $N$.
It is therefore plausible to assume that $M^2 < N \beta$ and so,
the procedure of calculating the above matrix entries comes at a rather low price: 
It requires less than $O(N \beta)$ operations.

In practice, finding the kernel (i.e., the subspace spanned by the vectors representing conserved quantities) of the equal-time covariance matrix $\bEta$ numerically may turn out to be a rather difficult task
especially for large system sizes. This is because of the statistical errors associated with the measured matrix elements of $\bEta$ which may 
eventually lead to {\it negative} eigenvalues, despite the fact that $\bEta$ 
should be strictly positive definite. The existence of negative eigenvalues implies 
that the errors and corresponding negative eigenvalues are comparable in size. Existence of negative eigenvalues
also has harsh consequences as far as the maximization procedure detailed in the previous section is concerned, as the process requires
that $\bEta$ be a positive-definite matrix. 
Therefore, a practical resolution of the above difficulties would be 
to simply add the eigenstates associated with the negative eigenvalues to the set of vectors $\{ \bb^{(k)} \}$ 
thereby ensuring that the subspace spanned by the remaining eigenstates of $\bEta$ has strictly positive eigenvalues
as required by the maximization process. As we shall later see,
this solution works very well in practice. 

Moving on to the second phase of the simulation, the operator $\hat{O}$ is constructed using the optimized set of parameters
whose values were set at the end of the first phase of the simulation, and its time correlations are measured. 
The construction of the composite operator $\hat{O}$ consists of $O(N \beta)$ operations 
corresponding to its evaluation ($M \sim N$ operations) at each ``time-slice'' of which there are of the order of $\beta$.  Since the time correlations of only one operator are being measured, calculation of the actual correlation data requires of the order of $2\beta \log \beta$ operations if one uses 
the Fast Fourier Transform algorithm to compute them.
This procedure may therefore be considered cheap computationally as well. 

Since the measurement of the optimized correlation function requires two phases 
of the simulation, it may seem that it also demands more computation time 
as compared to correlation function measurements 
that do not require optimization. 
In practice however, one finds that this is not the case. When using an optimized correlation function, two effects
come into play. First, the slowest-decaying exponent of the correlation function
becomes dominant at much shorter imaginary times and second, the correlation function being `lifted' above the horizontal axis 
results in a much less noisy curve. In practice, as we shall see in the next section, these effects 
translate to an overall effect of a much shorter needed computation time.  

\subsection{The two-operator case}
The simplest nontrivial example for a somewhat-optimized measurable operator one can think of and which can be given a closed-form expression, is one in which
the operator to be optimized is a combination of only two other operators, namely, $\hat{O}=\alpha_1 \hat{O}_1 + \alpha_2 \hat{O}_2$, 
where $\hat{O}_1$ and $\hat{O}_2$ correspond to two measurable quantities,
each corresponding to some linear combination of the basic operators. Assuming also that
the Hamiltonian is a linear combination of the two operators, namely $\hat{H} = \beta_1 \hat{O}_1 + \beta_2 \hat{O}_2$,
the condition $\langle \ba | \bb \rangle=0$, along with the normalization condition of $\hat{O}$ and the fact that the vector $(\beta_1,\beta_2)$ spans the kernel 
the associated $2 \times 2$ matrix $\bEta$, yields the answer:
\beq
\label{eq:partOpt}
(\alpha_1,\alpha_2) = \sqrt{\frac{-\beta_1 \beta_2}{\eta_{12}}} \frac{(-\beta_2,\beta_1)}{\beta_1^2 +\beta_2^2} \,,
\eeq
where no maximization of $\langle \ba | \bchi | \ba \rangle$ is needed.
In this case, the simulation need not be split into two phases
in order to determine the optimal coefficients.

\section{\label{sec:example}Illustrative examples}

In what follows, we demonstrate the effectiveness of using an optimized imaginary-time correlation function to extract the excitation gap. 
We do this by considering several problems taken from the field of Quantum Adiabatic Computation, 
in the context of which the Quantum Adiabatic Algorithm (QAA) \cite{farhi:01} has been devised to solve hard optimization problems efficiently on a quantum computer.

Within the framework of this approach, the efficiency, or complexity, of the QAA for a given input problem is often studied by analyzing the behavior of the excitation gap
of a one-parametric family of Hamiltonians that forms a linear interpolation between an easily solvable transverse-field Hamiltonian 
$\hat{H}_d$ (commonly referred to as a `driver' Hamiltonian) 
and a diagonal `problem' Hamiltonian $\hat{H}_p$ whose ground state encodes the solution to the optimization problem.
Put explicitly, the linear interpolation is
\begin{equation}
\hat{H}(s)=s \hat{H}_p +(1-s) \hat{H}_d \,,
\end{equation}
where $s \in [0,1]$. In these problems, the gap usually needs to be calculated for
several values of $s$ where the objective is to find the minimal gap among these (the reader is referred to Refs.~\cite{farhi:01,HY,YKS2008,YKS2010} for a more detailed description of the process). 

Here, we shall illustrate the advantages that come with using the optimization method described in previous sections by considering several typical instances
of a specific optimization problem of the ``constraint satisfaction'' type known 
as 1-in-3SAT (for a description of the problem see, e.g., Refs.~\cite{HY}  and~\cite{sat}), 
in which the Hamiltonian is a sum of $L$ three-local Hamiltonians, $\hat{H}_p=\sum_{a=1}^{L} \hat{H}_a$, where each term 
in the sum involves three spins picked randomly
from a pool of $N$ spins. Each local Hamiltonian $\hat{H}_a$ is given in this problem by the expression: 
\begin{eqnarray}
\hat{H}_{a} &=& \frac1{8}\Big(5-\sigma_{a_1}^z - \sigma_{a_2}^z - \sigma_{a_3}^z  \\\nonumber 
&+& \sigma_{a_1}^z \sigma_{a_2}^z +\sigma_{a_2}^z \sigma_{a_3}^z  +\sigma_{a_3}^z \sigma_{a_1}^z  
+3 \sigma_{a_1}^z \sigma_{a_2}^z \sigma_{a_3}^z \Big) \,,
\end{eqnarray}
where $a_i$ for $i=1,2,3$ label the participating spins and $\sigma_i^z$ is the $z$-component Pauli matrix acting on spin $i$.
The second part of the Hamiltonian is the driver Hamiltonian $\hat{H}_d$. It is a simple transverse-field Hamiltonian and is given by:
\begin{equation}
\hat{H}_d = -\frac1{2} \sum_{i=1}^{N}  \sigma_i^x  \,,
\end{equation}
where $\sigma_i^x$ is the $x$-component Pauli matrix acting on spin $i$.
\subsection{A 16-spin system}
Here, we analyze an instance of the 1-in-3SAT problem in which the number of spins in the system is $N=16$ and the number of clauses, each involving a 
randomly-chosen triplet of spins, is $L=13$. 
The relatively small size of the system will allow us to compare the QMC-based extracted gap against the corresponding exact-diagonalization result.
The chosen inverse temperature is $\beta=128$  (which, as we shall see obeys $\beta \Delta E_1 \gg 1$) and the value chosen for the parameter $s$ in
this example is $s=0.5$, which puts it very close to the location of the minimum gap. 

To illustrate the effectiveness of using an optimized correlation function, we will also compare our results against those obtained from 
a `partially-optimized' correlation function based on an optimization with respect to only two coefficients, and a `non-optimized' correlation function
whose coefficients are not optimized by any means but are chosen randomly instead.

The specific QMC method we use here to measure the excitation gap is known as the stochastic series
expansion (SSE) algorithm \cite{SSE1,SSE2}. This method involves a
Taylor series expansion of the partition function $Z=\Tr[\e^{-\beta \hat{H}}]$
and uses a discrete representation of continuous imaginary time.
Similarly to current path integral formulations in continuous imaginary time \cite{contQMC1,contQMC2,contQMC3}, this discretization
does not introduce errors into the algorithm.

Within the SSE scheme applied to the problem at hand, the `natural' operators to measure are 
the $N$ non-diagonal operators $\hat{A}_i=-\frac1{2} (1-s) \sigma_i^x$ with $i=1..N$ and the $L$ 
diagonal operators $\hat{A}_i=s \hat{H}_{a}$ with $a=1..L$ where $i=N+a$.
In the current example this amounts to a total of $M=N+L=29$ basic operators. 
The to-be-optimized operator $\hat{O}$ will thus be a linear combination of those. 
Associated with the model studied here is only one conserved quantity --
the energy of the system (hence the number of conserved quantities is $N_c=1$). 
It corresponds to the set of coefficients $\beta^{(1)}_i=1$ (with $i=1..M$). 

To obtain the optimal operator $\hat{O}$, we calculated in the course of the simulation the matrix entries $\chi_{ij}$ and $\eta_{ij}$
each corresponding to an easily measured quantity [see Eqs.~(\ref{eq:chiij})
and~(\ref{eq:etaij})]. The expressions for the various matrix
elements as they are implemented within the SSE algorithm are derived in Appendix~\ref{app:sseImp} for the convenience of the reader.

\begin{figure}
\begin{center}
\includegraphics[width=\figurewidth]{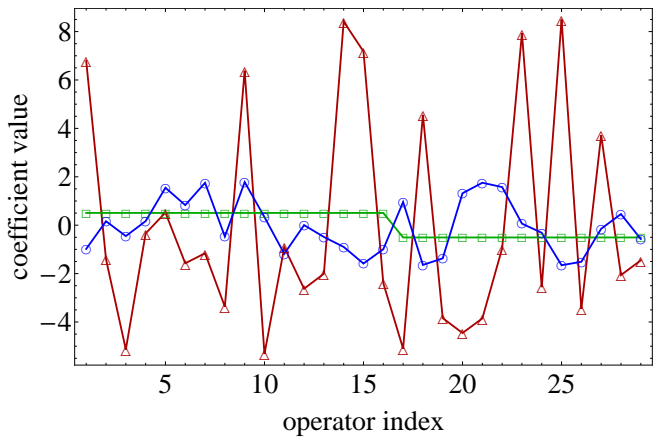} 
\caption{(Color online) Optimized (red triangles), `partially-optimized' (green squares) and `non-optimized' (i.e., a randomly-generated operator - blue circles) coefficients values of the operator $\hat{O}$ whose different-time correlations are
used to extract the gap. The values shown are normalized according to $||\hat{O}||=1$
or equivalently $\langle \ba | \bEta | \ba \rangle=1$. 
The horizontal line indicates the indices of the basic operators as they are defined in the text.}
\label{fig:bondVals}\vspace{-0.7cm}
\end{center}
\end{figure}

\begin{figure}[htp]
\begin{center}
\includegraphics[width=\figurewidth]{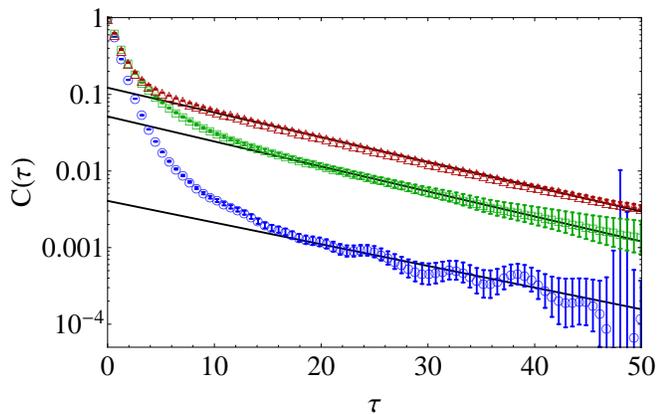}
\caption{(Color online) Optimized (red triangles), `partially-optimized' (green squares) and `non-optimized' (blue circles) correlation functions as functions of imaginary time on a log-linear scale for an instance of a 1-in-3SAT problem with $N=16$ 
interacting spin-$1/2$ particles. As the figure indicates, the optimization process substantially reduces the contributions from higher-order coefficients leaving 
even in relatively short times only the slowest-decaying 
exponent. The resulting optimal correlation function was easily fit with a linear function at the appropriate region producing an accurate prediction of the excitation gap
(the various linear fits are the solid lines). The partially-optimized correlation function is also a vast improvement over the non-optimized correlation function which in turn yields results of a much poorer quality. 
For the latter correlation function, 
there seems to be no region where a linear behavior is obvious, as higher-order contributions are evident throughout the (also very noisy) examined region.}
\label{fig:main}\vspace{-0.7cm}
\end{center}
\end{figure}

As a next step, we followed the procedure outlined in Sec.~\ref{sec:optProc} and
diagonalized the matrix $\bEta$ in order to find its kernel. As expected
from positive semi-definite matrices, the resulting eigenvalues 
were all found to be positive. One particularly small eigenvalue was also found and was immediately identified as the `zero' eigenvalue corresponding to the one conserved quantity
-- the energy. Numerically, the eigenvalue was found to be
of the order of $10^{-7}$, whereas for comparison, the next smallest eigenvalue turned out to be about $10^{3}$ times greater.
We also found as expected, that both matrices shared the same kernel corresponding the eigenvector $(1,1,\ldots,1)$ which represents the Hamiltonian of the system.
After projecting out the kernel of $\bEta$ from the two matrices (i.e., reducing the matrices to $\bar{\bEta}$ and $\bar{\bchi}$), we calculated the coefficients $\{ \alpha_i \}$ by maximizing the expression given in Eq.~(\ref{eq:maxExpression}) to obtain the optimal operator $\hat{O}$. These coefficients are plotted in Fig.~\ref{fig:bondVals} (marked by triangles). 
As the figure indicates, the various coefficients $\alpha_i$ turn out to be quite different from one another,
revealing the nontrivial complexity of the optimal operator. 
The correlation function itself is plotted in Fig.~\ref{fig:main} accompanied by a linear fit with which the excitation gap is extracted.

For comparison, we have also considered a much simpler set of operators, namely $O_1=(1-s) \hat{H}_d =\sum_{i=1}^{N} \hat{A}_i$ and $\hat{O}_2=s \hat{H}_p =\sum_{i=N+1}^{M} \hat{A}_i$, to construct the operator 
$\hat{O}$ from. 
For this choice of operators
the coefficients of the Hamiltonian correspond to $\beta^{(1)}_1=\beta^{(1)}_2=1$. Plugging these into Eq.~(\ref{eq:partOpt}),
we end up with $\alpha_1=-\alpha_2$ and the (unnormalized)
`partially-optimized' operator is:
\beq \label{eq:opo}
\hat{O}_{\textrm p.o.}=s \hat{H}_p-(1-s) \hat{H}_d \,.
\eeq
Measurement of the correlation function corresponding to this operator is also depicted in Fig.~\ref{fig:main} (square data points),
and the corresponding coefficients are plotted in Fig.~\ref{fig:bondVals} for comparison.

Figs.~\ref{fig:bondVals} and~\ref{fig:main} also show the (normalized) coefficients and correlation function of a 
'non-optimized' operator constructed by
random assignments of the various coefficients taken from a uniform distribution
in the range $[-1,1]$ (these are marked by circles in the two figures).

A side-by-side comparison of the three correlation functions plotted in Fig.~\ref{fig:main} 
clearly indicates that while the non-optimized correlation function is very noisy and simply 
does not allow for any serious calculation of the gap at least for this chosen running time of the simulation, 
the two other optimized correlation functions are much less noisy: both exhibit a straight-line behavior 
at some point in imaginary time and enable the extraction of the gap. That being said, it is evident from the figure that  the dominance of the slowest-decaying 
exponent is apparent in the fully-optimized correlation function at much shorter times than it is for the partially-optimized 
correlation function. Moreover, the fully-optimized correlation function also has a superior signal-to-noise ratio at all times.

Calculation of the excitation gap by the linear fits to the logarithm of the three (fully-optimized, partially-optimized and non-optimized)
correlation functions yields the results summarized in Table~\ref{tab:t1}.
As the table indicates, for the parameters chosen here, a linear fit of the fully-optimized correlation function clearly yields
the most accurate result among the three, with a very small error (of $0.0006$) and the exact-diagonalization value falling well within the error bar of the extracted value ($0.38$ 
standard deviations).
The partially-optimized correlation function yields slightly poorer results (an error of $0.005$ and with the exact value being $0.31$ standard deviations away),
although one could argue that such `partial' optimization might be enough for certain purposes. 
In this case, with essentially only one adjustable parameter, the partially-optimized correlation function is 
only slightly worse than the fully-optimized one. 
The non-optimized data on the other hand produced
much larger errors ($0.02$) and a rather poor fit, as one would expect from such noisy correlation function.

\begin{table}
\begin{center}
\begin{tabular}{||c|c|c||}
\hline\hline
  & Gap value and its error& Deviation from \\
 Operator-type  &  (the relative error is &  ED (in standard \\
   & shown in parentheses)  &  deviations) \\
\hline
fully-optimized  & $0.0747 \pm 0.0006 \quad (0.8 \%)$ & $0.38$  \\
partially-optimized  & $0.076 \pm 0.005 \quad (6.6 \%)$ & $0.31$ \\
non-optimized  & $0.053 \pm 0.02 \quad (38 \%)$ & $1.07$  \\
\hline\hline
\end{tabular}
\end{center}
\caption{
Calculated excitation gaps for the $16$-spin system as they were extracted from the three tested correlation functions, and their deviation from the exact-diagonalization (ED) value measured in number of standard deviations. 
As the table indicates, the optimized correlation function yields a much better prediction of the gap than the partially- and non-optimized correlation functions.
The exact diagonalization value is $\Delta E_1= 0.07447$ and the chosen inverse-temperature here is $\beta = 128$.
The requirement that the system be in its ground state is therefore satisfied as $\beta \Delta E \approx 10$. 
The errors reported here were determined by standard least-squares fitting analysis on the correlation-function data.
}
\vspace{-0.5cm}
\label{tab:t1}
\end{table}

\subsection{Larger spin systems}

In what follows, we present the results of an analysis similar to the one performed in the previous section 
but applied to larger spin systems. For these, the expected gap is much smaller and is also naturally more difficult to extract.  Here we shall consider two random
instances of the 1-in-3SAT type 
corresponding to systems with $N=48$ and $N=64$ spins, and with $L=38$ and $L=51$ clauses, respectively. 
While in these examples a comparison with exact-diagonalization results is unavailable,
it is nonetheless advantageous to make a comparison
between the fully-optimized, partially-optimized and non-optimized correlation functions.

\begin{figure}
\begin{center}
\includegraphics[width=\figurewidth]{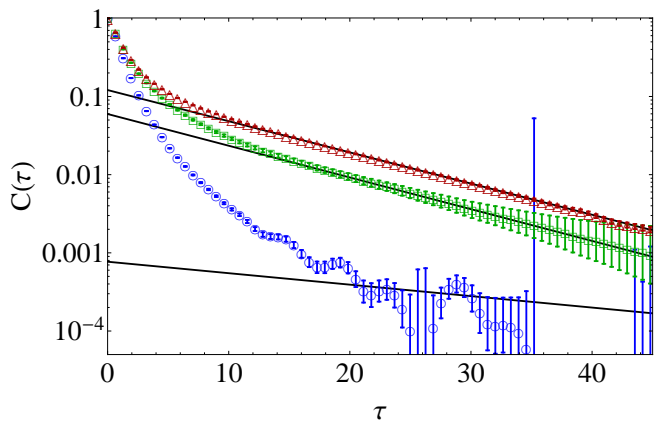}
\caption{(Color online) Optimized (red triangles), `partially-optimized' (green squares) and `non-optimized' (blue circles) correlation functions as functions of imaginary time on a log-linear scale for an instance of the 1-in-3SAT problem with $N=48$ interacting spin-1/2 particles. The figure shows that the optimal correlation function is easily fit with a linear function at the appropriate region producing an accurate prediction of the excitation gap
(the various linear fits are the solid lines). The partially-optimized correlation function is also a vast improvement over the non-optimized correlation function which in turn yields results of a much poorer quality. }
\label{fig:main48}\vspace{-0.7cm}
\end{center}
\end{figure}

The correlation functions obtained for the $N=48$ system 
(the adiabatic parameter $s$ and inverse temperature $\beta$ were chosen to be $s=0.5$ and $\beta=256$) are shown in Fig.~\ref{fig:main48} along with corresponding linear fits
from which the gaps are then extracted. The optimized correlation function was obtained via 
the maximization of the integrated susceptibility with respect to all of the $M=48+38=86$ coefficients $\alpha_i$. 
In this specific example, the subspace spanned by five eigenvectors corresponding to five negative eigenvalues 
that were found at the end of the first phase of the simulation have been discarded.
The second correlation function is the partially-optimized one corresponding to the
operator $\hat{O}_{\textrm p.o.}$ given in Eq.~(\ref{eq:opo}), and the third correlation function
tested here is the non-optimized one, based on randomly-assigned coefficients.

As in the previous example, it is evident from Fig.~\ref{fig:main48} that the fully-optimized correlation function is much less noisy and therefore 
produces a much more accurate reading of the gap with the smallest uncertainty $\Delta E_1=0.065 \pm 0.0015$ ($2.3 \%$ relative error). For comparison, the partially-optimized
correlation function yielded $\Delta E_1=0.064 \pm 0.005$ ($7.8 \%$ relative error). The non-optimized correlation functions 
turned out to have very large statistical errors and is therefore completely unreliable for the extraction of the gap.   
These results are summarized in Table \ref{tab:t2}. 

\begin{table}
\begin{center}
\begin{tabular}{||c|c|c|c||}
\hline\hline
\begin{tabular}{c} System \\size \end{tabular} & Operator-type  & \begin{tabular}{c} Gap value \\and its error \end{tabular} & 
\begin{tabular}{c} Relative \\error \end{tabular}\\
\hline \hline
$N=48$ &  
\begin{tabular}{c} fully-optimized \\partially-optimized \end{tabular}
 & \begin{tabular}{c} $0.065 \pm 0.0015$ \\$0.064 \pm 0.005$ \end{tabular}  & 
\begin{tabular}{c} $2.3\%$ \\$7.8\%$ \end{tabular}
 \\
\hline
$N=64$ &  
\begin{tabular}{c} fully-optimized \\partially-optimized \end{tabular}
 & \begin{tabular}{c} $0.051 \pm 0.001$ \\$0.052 \pm 0.007$ \end{tabular}  & 
\begin{tabular}{c} $2\%$ \\$13\%$ \end{tabular} \\
\hline\hline
\end{tabular}
\end{center}
\caption{
Calculated excitation gaps for the $48$- and $64$-spin systems as they were extracted from the three tested correlation functions and their relative errors. 
As the table indicates, the optimized correlation function yields a much more accurate prediction of the gap than the partially- and non-optimized correlation functions
for both sizes.
}
\vspace{-0.5cm}
\label{tab:t2}
\end{table}

\begin{figure}
\begin{center}
\includegraphics[width=\figurewidth]{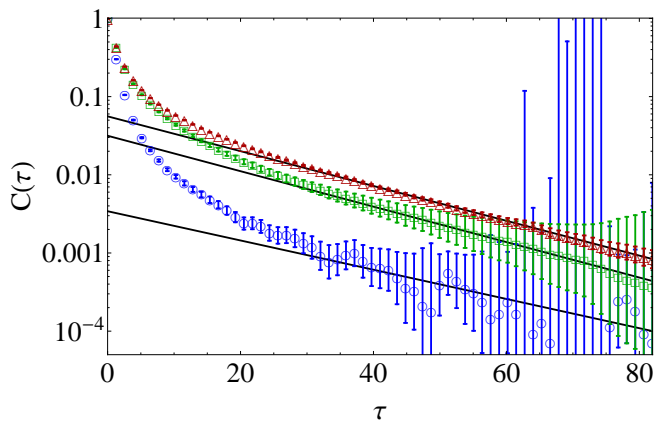}
\caption{(Color online) Same as Fig.~\ref{fig:main48} but for a system with $N=64$ spins.
Here too the optimal correlation function (red triangles) is much less noisy than the other partially-optimized (green squares)
and non-optimized (blue circles) correlation functions, eventually leading to a more accurate prediction of the system gap.
Of the two latter correlation functions, the non-optimized one is particularly noisy in this example and yields an extremely poor estimation of the gap.}
\label{fig:main64}\vspace{-0.7cm}
\end{center}
\end{figure}

Similarly to the $N=48$ case analyzed above, we present in Fig.~\ref{fig:main64} the correlation functions obtained in the analysis of a $64$-spin system with $51$ clauses
and a total of $M=115$ real coefficients (here, $s=0.52$ and $\beta=256$).
In this example, a $23$-dimensional subspace has been discarded in the maximization procedure, leaving a $92$-dimensional parameter-space with respect 
to which maximization is performed. 
As in the previous examples, here too the non-optimized, i.e., random, correlation function is practically useless for obtaining the gap,
whereas the partially-optimized and fully-optimized correlation functions
are much more suitable for the task. 

As expected, the fully-optimized correlation
function is significantly less noisy than the partially-optimized
one giving a much more accurate value for the gap:
$\Delta E_1=0.051 \pm 0.001$ ($2 \%$ relative error) for the fully-optimized versus $\Delta E_1=0.052 \pm 0.007$ ($13 \%$ relative error) for the partially-optimized one.  
Again, the non-optimized correlation function is much too noisy for any reliable reading of the gap
(see also Table \ref{tab:t2}). 

\section{\label{sec:sum}Summary and conclusions}
In this paper, we demonstrated the effectiveness of calculating and utilizing 
optimized imaginary-time correlation functions toward 
the extraction of excited-state information within quantum Monte Carlo simulations.
We have given a prescription for optimizing the operator whose imaginary-time correlation function would be best suited for gap calculations.
The optimization is based on maximizing the integrated susceptibility of the operator whose time-correlations are to be measured.
We have also illustrated the benefits associated with evaluating optimized correlation function 
compared to other non-optimized functions, and 
confirmed numerically that determining the gap to the first excited state from optimized functions 
is considerably more accurate than corresponding results obtained from non-optimized correlation functions. We have also commented on the relatively low cost of the procedure.

While in this study the main focus was on the extraction of the gap to the first excited state, 
it should be noted that the optimization procedure
whose derivation was presented here may of course be subjected to other more sophisticated methods aimed at obtaining a fuller picture 
of the spectrum of the Hamiltonian of the system, i.e., energies of more excited states (see e.g., Ref.~\cite{maxEnt,maxEnt2,maxEnt3}). These methods
of analysis can readily be applied to 
the correlation function obtained in the process described 
here.

The results of the analysis performed on the exemplary problems presented in the Sec.~\ref{sec:example} show that 
optimization of the correlation function results in both smaller statistical errors of the correlation function and also
a dominance of the slowest-decaying exponent at shorter imaginary times. This implies that
in order to obtain decent results for the gap from non-optimized correlation functions, 
a substantially longer running time of the simulation is needed (and perhaps also a larger value of the inverse-temperature $\beta$).
The results therefore establish the importance of applying
at least some optimization to measured correlation functions, as this may have
considerable effects on the ability to extract the gap and on the accuracy of the obtained value. 
We therefore expect that the procedure presented in this manuscript will be useful wherever gap calculations are needed numerically.

\begin{acknowledgments}
We thank Peter Young, Victor Martin-Mayor, Nir Lev and Anders Sandvik for useful comments and discussions.  
This work is supported in part by the National
Security Agency (NSA) under Army Research Office (ARO) contract number
W911NF-09-1-0391, and in part by the National Science Foundation under Grant
No.~DMR-0906366.
\end{acknowledgments}

\appendix

\section{\label{app:sseImp} Stochastic Series Expansion (SSE) measurements of the matrices $\bEta$ and $\bchi$}
In what follows we derive the explicit expressions needed for the measurements of
the matrix entries $\eta_{ij}$ and $\chi_{ij}$ within the framework of the stochastic series expansion (SSE) algorithm. 
As discussed in Sec.~\ref{sec:ocf}, calculation of 
the optimal correlation function requires 
 the evaluation of the set of coefficients $\{ \alpha_i \}$ 
which is obtained by an algebraic manipulation of the 
matrix elements $\chi_{ij}$ and $\eta_{ij}$ as they are defined in the main text [see Eqs.~(\ref{eq:chiij}) and~(\ref{eq:etaij})]. 
These are expressed in terms of the `basic' operators of the QMC technique being used. 
Within the SSE, the basic operators that are most easily measured
are the so-called `bond' operators which we denote here by $\hat{A}_i$.
These are the local operators that comprise the Hamiltonian, namely:
\beq
\hat{H} = \sum_{i=1}^M \hat{A}_i \,,
\eeq
where each of the bond operators $\hat{A}_i$ acts only on a limited number of particles.
The reader is referred to Sec.~\ref{sec:example} for the bond operators 
of the 1-in-3SAT problem. 
Within the SSE, the expectation values of the bond operators are
obtained by simply counting their occurrences in the `operator sequence' 
that defines the instantaneous configuration of the system (for 
a more detailed discussion of the SSE technique, see, e.g., \cite{SSE1,SSE2}),
namely:
\beq
\langle \hat{A}_i \rangle =-\frac1{\beta} \langle N_i \rangle \,,
\eeq
where $N_i$ is the number of times the operator $\hat{A}_i$ appears in the sequence.
Similarly, it is easy to show that 
\beq
\int_0^{\beta} \rmd \tau \langle \hat{A}_{i_1}(\tau) \hat{A}_{i_2}(0)\rangle  = \frac1{\beta}
\big( \langle N_{i_1} N_{i_2}\rangle - \delta_{i_1,i_2} \langle N_{i_1} \rangle \big) \,,
\eeq 
where $i_1$ and $i_2$ are two arbitrary bond indices \cite{SSE1}.
From the above equation, it follows that the matrix elements of $\bchi$ are simply given by:
\beq
\chi_{ij} = \frac1{\beta} \left(
\langle N_{i_1} N_{i_2} \rangle- \delta_{i_1,i_2} \langle N_{i_1} \rangle - \langle N_{i_1}\rangle \langle  N_{i_2} \rangle \right)\,.
\eeq
In addition, expectation  values of products of bond operators are obtained
using:
\beq
\langle \hat{H}_{i_1} \hat{H}_{i_2} \rangle = \frac1{\beta^2} \langle (n-1) N_{i_{12}}\rangle \,,
\eeq
where $N_{i_{12}}$ denotes the number of ordered subsequences  $ \hat{A}_{i_1} \hat{A}_{i_2}$ in the operator sequence and  $n=\sum_{i=1}^{M} N_i$ is the total number of operators in the sequence. 
The matrix elements of $\bEta$ are therefore similarly given by the expression:
\beq
\eta_{ij} = \frac1{\beta^2} \left(
\frac1{2} \langle (n-1) (N_{i_{12}} +N_{i_{21}}) \rangle - \langle N_{i_1}\rangle \langle  N_{i_2} \rangle \right).
\eeq


\begin{thebibliography}{100}
\bibitem{qmcRef1}
D. P. Landau and K. Binder, {\it A guide to Monte Carlo simulations in statistical physics}, (Cambridge, Cambridge University Press, 2005).
\bibitem{qmcRef2}
M. E. J. Newman and G. T. Barkema, { \it Monte Carlo Methods in Statistical Physics}, (Oxford, Clarendon Press, 1999). 
\bibitem{DMRG1}
S.~R. White, Phys. Rev. Lett. {\bf 69}, 2863 (1992).
\bibitem{DMRG2}
K. Hallberg, Adv. Phys {\bf 55}, 477 (2006).
\bibitem{excitedRefPhys}
M. Takahashi, in {\it Quantum Monte Carlo Methods in Condensed Matter Physics}, Ed. M. Suzuki, (Singapore, World Scientific, 1993). 
\bibitem{massgap1}
B. Berg, A. Billoire and C. Rebbi, Annals of Physics {\bf 142}, 185 (1982).
\bibitem{massgap2}
G. Fox, R. Gupta, O. Martin and S. Otto, Nuclear Physics {\bf B205}, 188 (182). 
\bibitem{massgap3}
K. Langfeld, H. Reinhardt, and O. Tennert, Phys. Rev. D {\bf 56}, 6798 (1997). 
\bibitem{excitedNuc}
V. J. Emery and A. M. Sessler, ,Phys. Rev. {\bf 119}, 248 (1960). 
\bibitem{excitedNuc2}
M. C. Jain and Y. R. Waghmare, J. Phys. A: Gen. Phys. {\bf 3}, 274 (1970).
\bibitem{excitedRefChem}
W.~J. Hehre, {\it A Guide to Molecular Mechanics and Quantum Chemical Calculations} (Irvine, Wavefunction Inc., 2003).
\bibitem{excitedRefChem2}
D. M. Ceperley and B. Bernu, J. Chem. Phys. {\bf 89}, 6316 (1988). 
\bibitem{maxEnt}
D. Blume, M. Lewerenz, P. Niyaz and K. B. Whaley,
Phys. Rev. E {\bf 55}, 3664 (1997);
\bibitem{maxEnt2}
D. Blume, M. Lewerenz and K.B. Whaley, Mathematics and Computers in Simulation {\bf 47}, 133 (1998)  
\bibitem{maxEnt3}
P. Huang, A. Viel and K. B. Whaley, in {\it Recent Advances in Quantum Monte Carlo Methods Part II}, (World Scientific, Singapore, 2002). 
\bibitem{improved}
U. Wolff, Phys. Rev. Lett. {\bf 62}, 361 (1989).
\bibitem{smeared1}
S. G\"usken, U. L\"ow, K.~H. M\"utter, R. Sommer, A. Patel and K. Schilling, Phys. Lett. B {\bf 227}, 266 (1989).
\bibitem{smeared2}
C. Best, M. G\"ockeler, R. Horsley, E.~M. Ilgenfritz, H. Perlt, P. Rakow, A. Sch\"afer, G. Schierholz, A. Schiller, and S. Schramm,
Phys. Rev. D {\bf 56}, 2743 (1997). 
\bibitem{cont}
A. W. Sandvik and R. R. P. Singh, Phys. Rev. Lett. {\bf 86}, 528 (2001).
\bibitem{sandvik}
We thank Anders Sandvik for suggesting that the analysis 
of correlation functions be done along these lines.
Anders Sandvik, private communication. 
\bibitem{farhi:01}
E. Farhi, J. Goldstone, S. Gutmann, J. Lapan, A. Lundgren, and D. Preda, Science {\bf 292}, 472 (2001), a longer version of the paper appeared in arXiv:quant-ph/0104129.
\bibitem{HY}
I. Hen and A. P. Young, Phys. Rev. E {\bf 84}, 061152 (2011).
\bibitem{YKS2008}
A. P. Young, S. Knysh, and V. N. Smelyanskiy, Phys. Rev. Lett. {\bf 101}, 170503 (2008). 
\bibitem{YKS2010}
A. P. Young, S. Knysh, and V. N. Smelyanskiy, Phys. Rev. Lett. {\bf 104}, 020502 (2010). 
\bibitem{sat}
M. K. Garey and D. S. Johnson, {\it Computers and Intractability. A Guide to the Theory of NP-Completeness}, (H. H. Freeman, New York, 1997).
\bibitem{SSE1}
A. W. Sandvik, Phys. Rev. B {\bf 59}, R14157 (1999). 
\bibitem{SSE2}
A. W. Sandvik, J. Phys. A {\bf 25}, 3667 (1992). 
\bibitem{contQMC1} E. Farhi, J. Goldstone, D. Gosset, S. Gutmann, H. B. Meyer, and P. Shor, Quantum Information \& Computation {\bf 11}, 181 (2009), (arXiv:0909.4766). 
\bibitem{contQMC2} F. Krzakala, A. Rosso, G. Semerjian, and F. Zamponi, Phys. Rev. B {\bf 78}, 134428 (2008), (arXiv:0807.2553). 
\bibitem{contQMC3}
O. F. Syljuasen and A. W. Sandvik, Phys. Rev. E {\bf 66}, 046701 (2002). 
\end{thebibliography}
\end{document}